\documentclass[a4paper,11pt,oneside]{article}

\usepackage[english]{babel}
\usepackage[T1]{fontenc}
\usepackage[utf8]{inputenc}

\usepackage[pdftex,unicode]{hyperref}
\hypersetup{pdftitle=The Role of Shopping Mission in Retail Customer Segmentation}
\hypersetup{pdfauthor=Ondřej Sokol and Vladimír Holý}

\usepackage[margin=60pt]{geometry}
\usepackage{amsmath}
\usepackage{amssymb}
\usepackage{bm}

\usepackage[group-minimum-digits=3]{siunitx}

\usepackage{enumitem}

\usepackage{graphicx}
\usepackage{float}
\usepackage{hhline}
\usepackage{multirow}
\usepackage{rotating}
\usepackage{booktabs}

\usepackage[authoryear]{natbib}

\begin{document}

\begin{center}
{\Large \bfseries The Role of Shopping Mission in Retail Customer Segmentation}
\end{center}

\begin{center}
{\bfseries Ondřej Sokol} \\
University of Economics, Prague \\
Winston Churchill Square 4, 130 67 Prague 3, Czechia \\
\href{mailto:ondrej.sokol@vse.cz}{ondrej.sokol@vse.cz} \\
Corresponding Author
\end{center}

\begin{center}
{\bfseries Vladimír Holý} \\
University of Economics, Prague \\
Winston Churchill Square 4, 130 67 Prague 3, Czechia \\
\href{mailto:vladimir.holy@vse.cz}{vladimir.holy@vse.cz}
\end{center}

\begin{center}
{\itshape \today}
\end{center}

\noindent
\textbf{Abstract:}
In retailing, it is important to understand customer behavior and determine customer value. A useful tool to achieve such goals is the cluster analysis of transaction data. Typically, a customer segmentation is based on the recency, frequency and monetary value of shopping or the structure of purchased products. We take a different approach and base our segmentation on the shopping mission -- reason why a customer visits the shop. Shopping missions include focused purchases of specific product categories and general purchases of various sizes. In an application to a Czech drugstore chain, we show that the proposed segmentation brings unique information about customers and should be used alongside the traditional methods.
\\

\\

\\

\section{Introduction}
\label{sec:intro}

Retail chains have a huge amount of sales data available. An analysis of these data strives to understand  \emph{customer behavior} and determine  \emph{customer value} in order to increase profits. The information about customers can be utilized in various areas such as new product development \citep{Li2012}, product positioning \citep{Gruca2003}, cross-category dependence \citep{Hruschka1999, Russell2000, Leeflang2008}, product complements and substitutes determination \citep{Srivastava1981, Chib2002}, category management \citep{Duchessi2004}, promotions planning \citep{Trappey2009}, online marketing \citep{Chen2009}, targeted advertising \citep{Jonker2004, Zhang2007}, product recommendation \citep{Liu2005}, product association rules \citep{Weng2016} and stock optimization \citep{Borin1994}. One of the tools used to achieve these goals is the \textit{cluster analysis}. 

There are many applications of the cluster analysis in retail business. Products sold by the shop can be clustered according to their characteristics in order to find substitutes and complements \citep{Srivastava1981} or target market \citep{Zhang2007}. A product categorization based solely on customer shopping patterns was proposed by \cite{Holy2017}. Customers can be segmented according to their demographics and lifestyle or their shopping behavior. A popular approach is to segment customers based on the \emph{recency, frequency and monetary value (RFM)} of their shopping \citep{Kahan1998, Miglautsch2000, Yang2004, Chen2009, Khajvand2011, Putra2012, Peker2017, Boon2016}. Another approach is to segment customers based on the \emph{purchased products structure (PPS)} \citep{Russell1997, Manchanda1999, Andrews2002, Tsai2004}. \cite{Lingras2014} and \cite{Ammar2016} simultaneously clustered both products and customers. Overall, the cluster analysis brings useful insight into the customer behavior and helps in the decision-making process, especially when combined with business knowledge \citep{Seret2014}.

We deal with  \emph{customer segmentation} using data from receipts. The traditional RFM and PPS segmentations answer the questions:
\begin{itemize}
\item When was the last time customers visited the shop?
\item How often do customers visit the shop?
\item How much money do customers spend?
\item What product categories do customers buy?
\end{itemize}
In our analysis, we propose to segment customers based on their \textit{shopping mission (SM)}. The proposed segmentation provides background for these crucial business questions:
\begin{itemize}
\item What is the purpose of customer visits?
\item Do customers visit the shop because of a specific product category?
\item Do customers buy products in other shops?
\end{itemize}
The proposed approach brings a new insight into the structure of customers. As a result, it can be used in many marketing areas such as the promotion planing and shelf management.

\begin{figure}
\begin{center}
\includegraphics[width=0.7\textwidth]{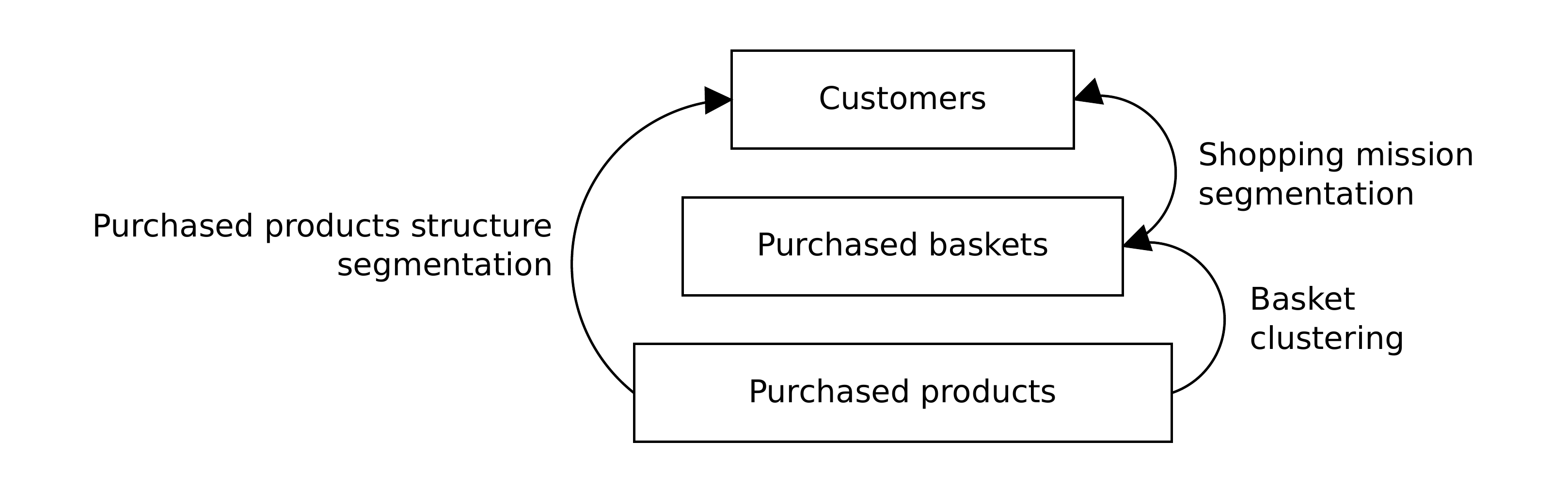} 
\caption{The process of the PPS segmentation and the SM segmentation of customers.}
\label{fig:segmentationTypes}
\end{center}
\end{figure}

The main idea of the proposed approach is as follows. We utilize the transaction data in  form of receipts which are linked to customers through the loyalty program. We first cluster individual baskets and then use this information to segment customers. This is illustrated in Figure \ref{fig:segmentationTypes} along with a comparison to the PPS segmentation. The method of \emph{k-means} is utilized for both the basket clustering and the customer segmentation. An analysis of a Czech drugstore chain shows that there are some customers who visit the shop with a \emph{focused purchase} of a specific product category in mind while others prefer a \emph{general purchase}. Another segmentation based on a shopping mission was presented by \cite{Reutterer2006}. The main difference compared to our proposed method is that we also consider the value of the basket. The proposed method is not meant to replace the RFM or PPS segmentation but rather to be used alongside and to bring a new perspective. The combination of the RFM, PPS and SM approaches forms a versatile segmentation based on a broad range of customer characteristics not tied to a single specific purpose. We emphasize the interpretability and usability by marketing departments and other experts involved in the retail decision-making process. 

\section{Transaction Data}
\label{sec:data}

We perform the analysis of retail business using \emph{transaction data}. The hierarchical structure of these data is illustrated in Figure \ref{fig:segmentationTypes}.

A \emph{product} is characterized by the brand, physical properties, and purpose. Note that the price of the product and whether the product is in sales promotion can vary over time and therefore we put it to the receipt data. Because there are many products, it is useful to aggregate them into \emph{product categories}. We denote the product categories as $K = \{K_i : i = 1,\ldots, n_K\}$. An individual product bought by the customer is referred to as the \emph{purchased product}. We denote the purchased products as $P = \{P_i : i = 1,\ldots, n_P\}$. Each purchased product belongs to a single product category.

A \emph{purchased basket} is a set of purchased products. We denote the purchased baskets as $B = \{B_i : i = 1,\ldots, n_B\}$. A specific purchased basket $B_i$ is a subset of all purchased products, i.e. $B_i \subset P$, $i = 1, \ldots, n_B$. A \emph{receipt} is a purchased basket with additional information about the customer ID, prices, sales promotion, date and time. 

A \emph{customer history} is a set of purchased baskets. We denote the customer history as $C = \{C_i : i = 1,\ldots, n_C\}$. A specific customer history $C_i$ is a subset of all purchased baskets, i.e. $C_i \subset B$, $i = 1, \ldots, n_C$. A \emph{customer} is a customer history with additional information about the contact, gender, age, number of children and other demographic information.

In the paper, we analyze a sample of real data. Our dataset consists of individual purchase data of one of the Czech drugstore retail chains. We use a three-month dataset of receipts which include more than $5.6$ million baskets bought by more than $1.5$ million customers with the loyalty card. Each row in a receipt stands for a single purchased product. The retail chain sells over 10 thousand products which are divided into 55 categories based on their purpose. This categorization was done by an expert.

\section{Segmentation Based on Recency, Frequency and Monetary Value}
\label{sec:rfm}

One of the most popular ways to segment customers is the clustering using data about \emph{recency, frequency and monetary value (RFM)} of their shopping. This method is fast and simple as its original purpose is to provide an easy-to-implement framework for quantifying customer behavior \citep{Kahan1998, Miglautsch2000}. 

\cite{Yang2004} described some shortcomings of RFM method (e.g. the inability of the RFM to generate the real differences among the RFM cells) and introduced a single predictor which is consolidated from the three variables of RFM. A method for the sequential pattern mining using the RFM segmentation was presented by \cite{Chen2009}. Another expansion of RFM by \cite{Peker2017} proposed to include customer relation length and periodicity in the customer segmentation.

The first RFM characteristic of a customer is the \emph{recency (Rec)}. Customers are segmented by the time of the last purchase occurrence. With the knowledge of the recency of the last purchase, the retailer can use different marketing techniques to attract customers who had been in the shop in the last week and customers who had not been there for months. In the broader concept, analysis of the occurrences of shopping in time can help for example to identify leaving customers, i.e. those who used to visit the shop frequently but their shopping behavior changed in the recent history. A company can use this knowledge to retain the customer using various marketing techniques. A customer $C_i$, $i = 1, \ldots, n_C$ is assigned to a recency segment $C^{Rec}_j$, $j = 1,\ldots,k_{Rec}$ according to the variable
\begin{equation}
IC_i^{Rec} = \min \{ \text{Days since a purchase by customer $i$} \}.
\end{equation}
The segments are either found by some clustering algorithm or defined by an expert.

The second RFM characteristic of a customer is the \emph{frequency (Frq)}. The purchase frequency is defined as the number of visits of a customer during a given time frame. Along with the average value of a basket, it is one of the most tracked key performance indicators. The marketing department can use the information about frequency and distinguish frequent customers from the ones who go to the shop just in case of emergency. While the goal for the loyal customers is to preserve their shopping behavior, the customers who rarely visit the shop should be recruited. A customer $C_i$, $i = 1, \ldots, n_C$ is assigned to a frequency segment $C^{Frq}_j$, $j = 1,\ldots,k_{Frq}$ according to the variable
\begin{equation}
IC_i^{Frq} = \frac{\text{Number of baskets purchased by customer $i$}}{\text{Time frame}}.
\end{equation}
As in the case of the recency, the segments are either found by some clustering algorithm or defined by an expert. 

The third RFM characteristic of a customer is the \emph{monetary value (Mon)}. A customer segmentation by monetary value can be done in various ways. A common approach is  to compute either the sums of all sales during a given time frame or the average value of baskets in a given time frame for each customer. The latter is used in a combination with the frequency analysis. Retailers can also focus on margins instead of sales. In our case, we assign a customer $C_i$, $i = 1, \ldots, n_C$  to a monetary value segment $C^{Mon}_j$, $j = 1,\ldots,k_{Mon}$ according to the variable
\begin{equation}
IC_i^{Mon} = \frac{\text{Total value of products purchased by customer $i$}}{\text{Time frame}}.
\end{equation}
Again, segments are either found by some clustering algorithm or defined by an expert. 

The combination of the above mentioned approaches forms the RFM segmentation. One approach to derive the RFM segmentation is to create the 3-dimensional matrix of all combinations of the $C^{Rec} = \{C^{Rec}_i : i = 1,\ldots,k_{Rec} \}$, $C^{Frq} = \{C^{Frq}_i : i = 1,\ldots,k_{Frq} \}$ and $C^{Mon} = \{C^{Mon}_i : i = 1,\ldots,k_{Mon} \}$ segmentations with size $k_{Rec} \times k_{Frq} \times k_{Mon}$. However, the number of clusters $k_{RFM} = k_{Rec} k_{Frq} k_{Mon}$ can be quite large. To reduce the number of segments, another approach may be used. For a given $k_{RFM}$, a customer $C_i$, $i = 1, \ldots, n_C$  can be assigned to a RFM segment $C^{RFM}_j$, $j = 1,\ldots,k_{RFM}$ according to the vector variable
\begin{equation}
IC_i^{RFM} = \left[ IC_i^{Rec}, IC_i^{Frq}, IC_i^{Mon} \right].
\end{equation}
The segments can be found by some clustering algorithms such as the k-means method. Despite its shortcomings, the RFM segmentation is commonly used across retail business for its simplicity and straightforward interpretation.

\section{Segmentation Based on Purchased Products Structure}
\label{sec:product}

Products in retail shops are often categorized based on their properties such as purpose, price, pack size and brand. The categorization of products can be done either by an expert or by an algorithm \citep{Srivastava1981, Zhang2007, Holy2017}. Subsequently, customers can be segmented using their receipts. We refer to this clustering as \emph{purchased product structure (PPS)} segmentation. The knowledge of their purchases is directly linked to product categories. Such analysis reveals commonly bought categories and therefore helps in targeting of marketing campaigns.

Segmentation of customers based on their category purchases was studied in \cite{Russell1997}, where authors segmented customers with respect to brand preference using household purchase data. Another approach of using product categorization on household data to analyze customer behavior was published by \cite{Manchanda1999}. A method for identifying customer segments with identical choice behaviors across product categories using logit model was presented by \cite{Andrews2002}. \cite{Tsai2004} dealt with clustering customers based on their purchase data linked to product categories and presented a methodology to ensure the quality of the resulting clustering. \cite{Lingras2014} and \cite{Ammar2016} utilized an iterative meta-clustering technique that uses clustering results from one set of objects to dynamically change the representation of another set of objects. The method is applied on product categorization and customer segmentation using supermarket basket data. 

We segment customers based on ratios of their purchases in each category. For a customer $C_i$, $i = 1, \ldots, n_C$, the PPS segmentation is based on information about a product category $j$, $j=1,\ldots,n_K$ given by
\begin{equation}
IC_i^{Cat,j} =  \frac{\text{Total value of products in $j$ purchased by $i$}}{\text{Total value of products purchased by $i$}}.
\end{equation}
A customer $C_i$, $i = 1, \ldots, n_C$ is then assigned to a PPS segment $C^{Frq}_j$, $j = 1,\ldots,k_{PPS}$ according to the vector variable
\begin{equation}
IC_i^{PPS} = \left[ IC_i^{Cat,1}, \ldots, IC_i^{Cat,n_K} \right].
\end{equation}
We find PPS segments using the k-means method. The optimal number of clusters $k_{PPS}$ is chosen according to the ratio of between cluster variance and total variance and Davies-Bouldin index alongside with a reasonable interpretation of resulting clusters.

\begin{figure}
\begin{center}
\includegraphics[width=0.7\textwidth]{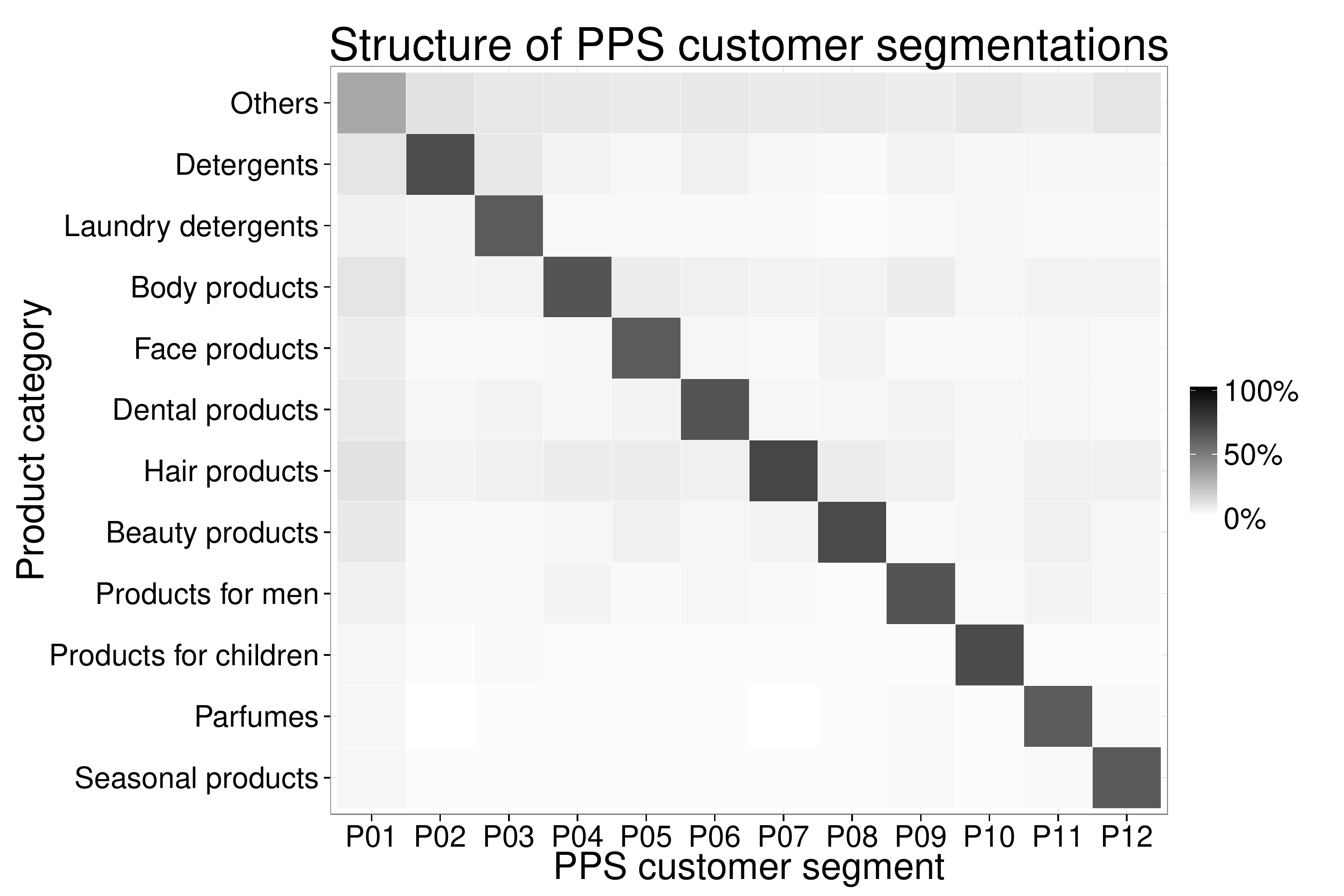} 
\caption{Ratios of product value from the PPS customer segments split into the product categories.}
\label{fig:productHeat}
\end{center}
\end{figure}

We perform a customer segmentation in a Czech drugstore chain according to 55 product categories. We find that in our case the optimal number of clusters is 12. Figure \ref{fig:productHeat} shows that there are 11 specialized segments and 1 general segment. The centers of specialized segments are composed of about 60 percent spendings in a single category. On the other hand, the general segment is fairly uniformly composed of over 15 popular categories. The distribution of customers assigned to the individual segments is alongside with labels of the dominant product categories shown in Table \ref{tab:productCount}. It is worth noting that despite having 55 categories, the top 15 categories comprise of over 98 percent of the total revenue.

\begin{table}
\begin{center}
\begin{tabular}{llr}
\toprule
Cluster & Description & Share \\
\midrule
P01 & General                              & 32\% \\
P02 & Specialized -- Detergents            &  6\% \\
P03 & Specialized -- Laundry detergents    &  6\% \\
P04 & Specialized -- Body products         &  6\% \\
P05 & Specialized -- Face products         &  6\% \\
P06 & Specialized -- Dental products       &  6\% \\
P07 & Specialized -- Hair products         &  8\% \\
P08 & Specialized -- Beauty products       & 10\% \\
P09 & Specialized -- Products for men      &  5\% \\
P10 & Specialized -- Products for children &  5\% \\
P11 & Specialized -- Perfumes              &  7\% \\
P12 & Specialized -- Seasonal products     &  4\% \\
\bottomrule
\end{tabular}
\caption{Distribution of the PPS customer segmentation.}
\label{tab:productCount}
\end{center}
\end{table}

We use the PPS segmentation as a base customer clustering according to categories they purchase. The next step is to compare it with a more complex approach featuring an intermediate step and adding the basket information to the segmentation. Figure \ref{fig:segmentationTypes} shows the process of both segmentations.

\section{Segmentation Based on Shopping Mission}
\label{sec:mission}

We propose an addition to the RFM and PPS segmentations. The above mentioned segmentations lack the information about the reason why customers visit the shop. Some customers visit the shop just to buy one product they need. This is called the \emph{focused purchase}. Other customers purchase large amounts of various products. This is called the \emph{general purchase}. Our goal is to design a reasonable segmentation reflecting the reasons why customers come to the shop, i.e. their \emph{shopping mission}. We use the term shopping mission in the sense of what products customers actually buy and in which type of basket. We can only guess the true motivation behind the visit (i.e. a customer is thirsty and therefore buys a bottle of water). From the marketing point of view, customers who come to the shop just for certain categories probably buy everything else in some other shop, therefore the goal is to transform them into regular customers. On the other hand, customers who already fulfill a majority of their needs in the shop are the most valuable ones and the goal of the marketing department is to retain them.

In the literature, the shopping mission or shopping motivation is often approached from a qualitative point of view. Hedonic shopping motivation and its effect in utilitarian enviroments was studied by \cite{Yim2014} using a field survey. Studies based on transaction data are present in the literature as well. \cite{Schroder2017} analyzed multi-category purchase decisions on the weekly basis using the item response theory models which allows to reveal characteristics of households for purchase decisions. Underlying latent activities of shoppers are also focus of the study by \cite{Hruschka2014} using topic models. Analysis of baskets using self-organizing maps was presented by \cite{Decker2003}. \cite{Reutterer2006} introduced a two-stage method of clustering customers using the basket clustering. In the first phase, baskets are clustered based on the purchased products. This is done using information whether the product appeared in the basket or not. In the second phase the customers are segmented based on their baskets. A method for identifying shopping mission using basket value and variety was proposed in \cite{Sarantopoulos2016}. 

\begin{figure}
\begin{center}
\includegraphics[width=0.7\textwidth]{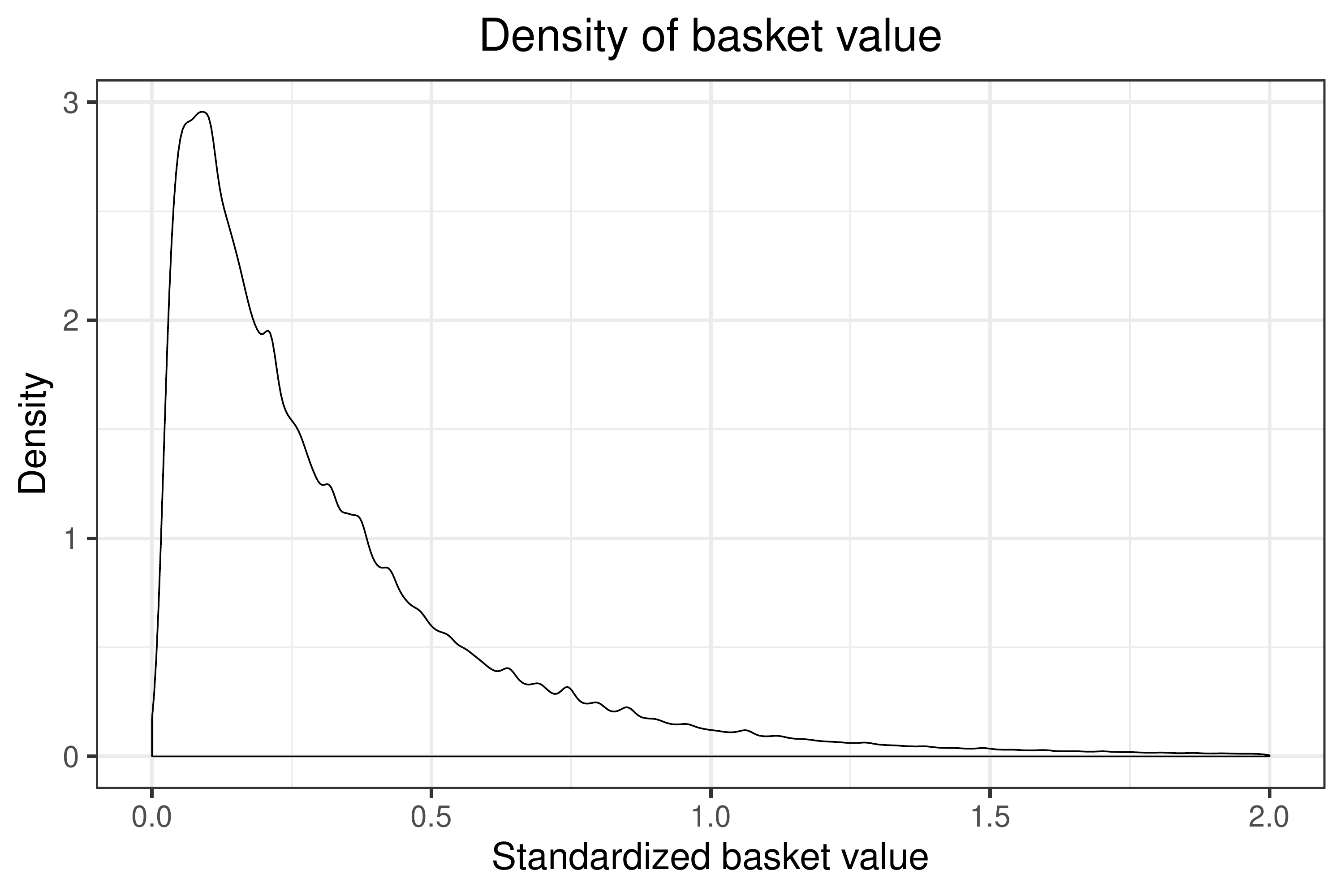} 
\caption{Estimated kernel density function of standardized basket value.}
\label{fig:revenue}
\end{center}
\end{figure}

For the clustering of baskets, we utilize purpose categories as well as the basket value resulting in easily interpretable clusters. In order to get reasonable clustering, the value of basket is standardized using the 95\% quantile of a basket value while the baskets with the value over this quantile are set to 1 due to a skewed distribution of the basket value. See Figure \ref{fig:revenue} for the kernel density function of the standardized basket value. In the clustering, each basket is then represented by a vector of non-negative ratios with unit sum and a standardized basket value ranging from 0 to 1. The interpretation is that we give similar weights to both the structure and the value of the basket. For basket $B_i$, $i = 1, \ldots, n_B$, the PPS clustering is based on information about product category $j$, $j=1,\ldots,n_K$ given by
\begin{equation}
IB_i^{Cat,j} =  \frac{\text{Total value of products in category $j$ in basket $i$}}{\text{Total sales of products in basket $i$}}
\end{equation}
and the information about value given by
\begin{equation}
IB_i^{Val} =  \min \left\{ \frac{\text{Total value of products in basket $i$}}{q_{95}}, 1 \right\},
\end{equation}
where  $q_{95}$ is the 95\% quantile of all basket values. A basket $B_i$, $i = 1, \ldots, n_B$ is then assigned to a SM cluster $B^{PPS}_j$, $j = 1,\ldots,k_{B}$ according to the vector variable
\begin{equation}
IB_i^{SM} = \left[ IB_i^{Cat,1}, \ldots, IB_i^{Cat,n_K}, IB_i^{Val} \right].
\end{equation}
We find the SM basket segments using the k-means method and choose the optimal number of clusters $k_{B}$ according to the ratio of between cluster variance and total variance and Davies-Bouldin index.

\begin{figure}
\begin{center}
\includegraphics[width=0.7\textwidth]{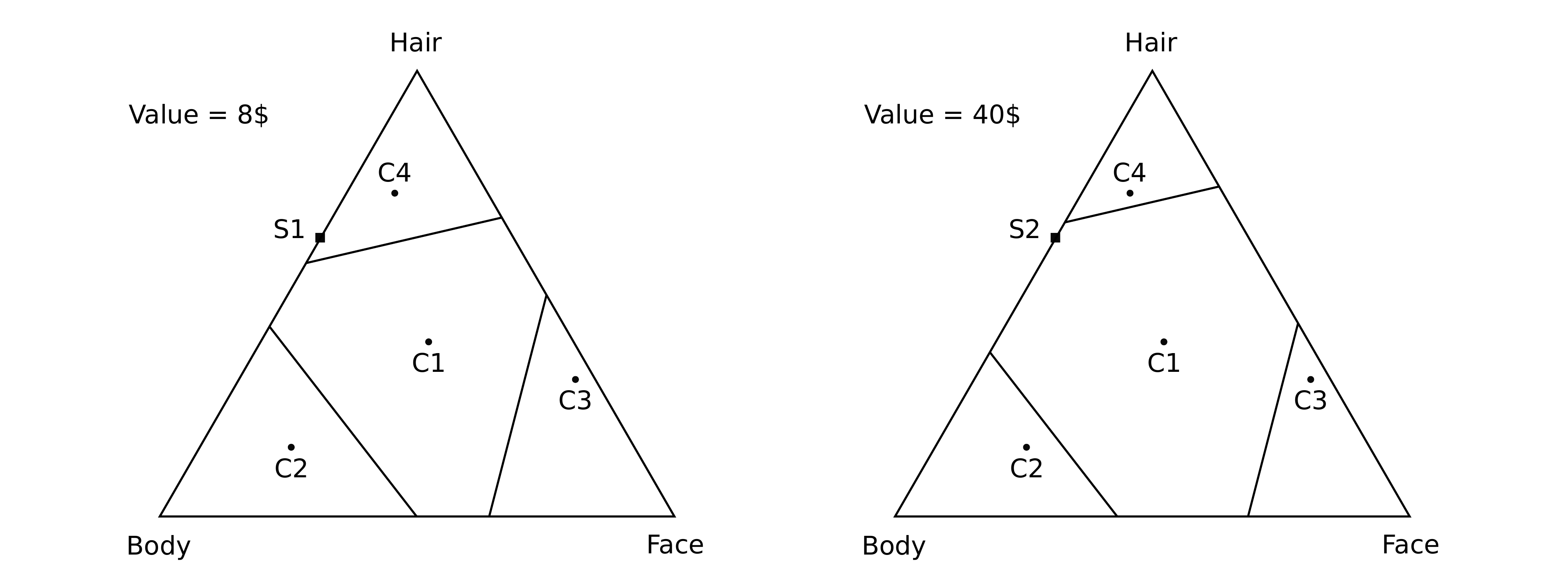} 
\caption{Illustration of the SM segmentation for different basket value levels.}
\label{fig:simplex}
\end{center}
\end{figure}

Our basket segmentation has the following geometric interpretation. Let us denote
\begin{equation}
IB_i^{Cat} = \left[ IB_i^{Cat,1}, \ldots, IB_i^{Cat,n_K} \right].
\end{equation}
For a given basket $i$, $IB_i^{Cat}$ then represents a point in a simplex of dimension $n_K$ while $IB_i^{Val}$ adds a depth to this simplex. The clustering is simply a dissectioning of this space. For simplicity, we focus on a low dimension of three product categories. The ratio of spending in a product category is then represented by a point in a triangle, whose vertices represent the exclusivity of a category in the basket. The value of basket adds depth to the triangle and forms a prism. Each basket is a point in this space. The centers of resulting clusters are inside the prism as well. In Figure \ref{fig:simplex}, we show two cuts of the prism with 4 cluster centers and defined cluster area for low and high basket value. The center C1 represents a point with a higher value than the others. Therefore its cluster area in cuts by basket value expands with a higher value of the basket. The baskets are assigned to the nearest center using the standard Euclidean distance. Let us consider the following example. First, we denote S1 the basket with hair products worth of 5\$, body products worth of 3\$ and no face products. Second, we denote S2 the basket with hair products worth of 25\$, body products worth of 15\$ and no face products. Both baskets have the same ratio of the three product categories and therefore have the same position in the simplex. Their value, however, differs making their depth different as well. As a result, they are assigned to different clusters. Basket S1 is closest to the center C4 while basket S2 is closest to the center C1.

\begin{figure}
\begin{center}
\includegraphics[width=0.7\textwidth]{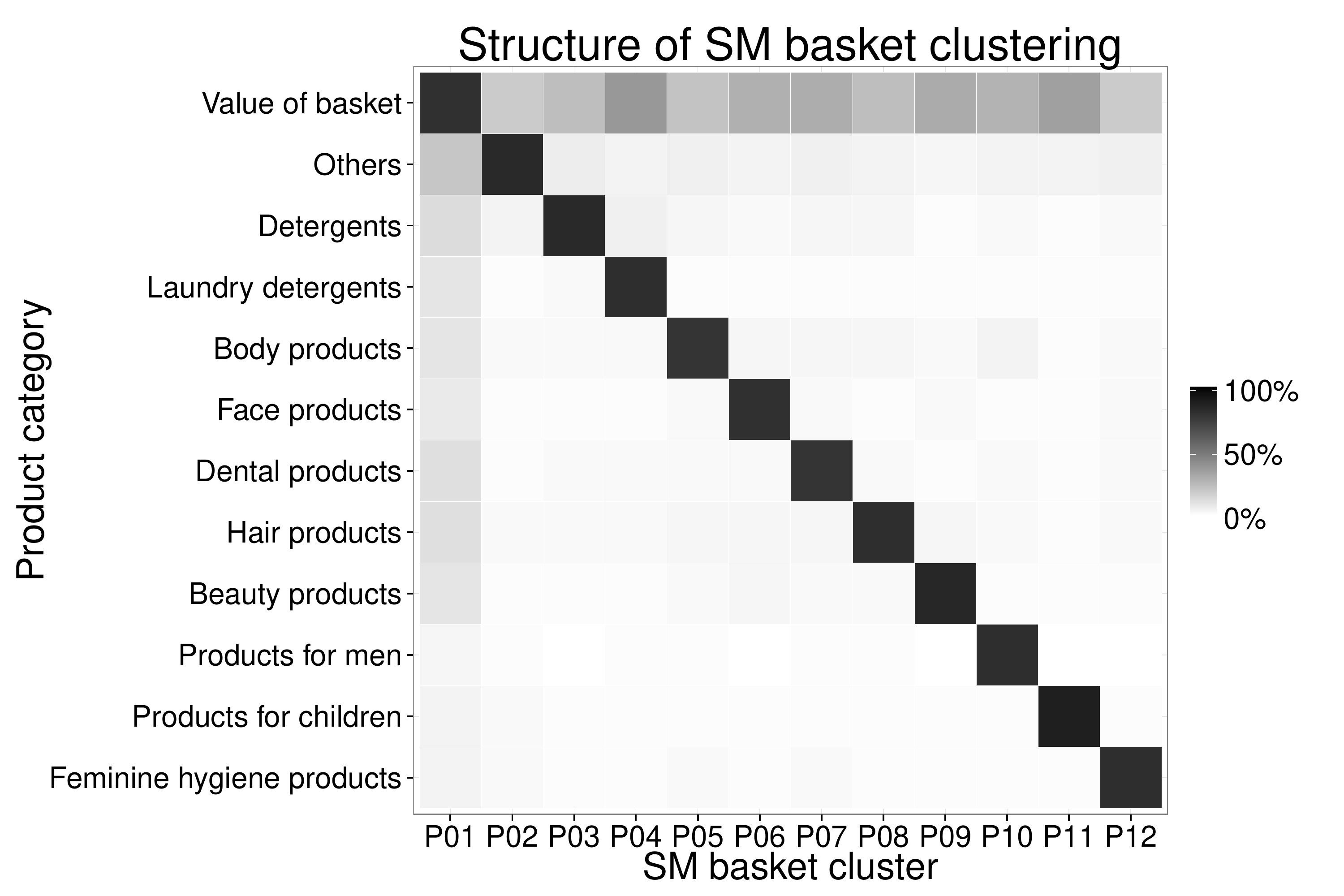} 
\caption{Ratios of product value from the SM basket clusters split into the product categories.}
\label{fig:basketHeat}
\end{center}
\end{figure}

We cluster baskets in a Czech drugstore chain according to 55 product categories. As expected the basket clusters are formed around previously mentioned well-selling categories. We find that 12 is the optimal number of clusters. Two clusters represent small and big universal baskets with no dominant category while the 10 others are focused on a single dominant category. The between cluster variance ratio is $0.78$ with 12 clusters while Davies-Bouldin index for 12 clusters has similar value to the other possible choices. The resulting distribution of baskets to the cluster as well as the interpretation of each cluster is shown in Table \ref{tab:basketCount}. Each cluster is named after the dominant category similarly to the PPS segmentation. The structure of each basket cluster (archetypes) is shown in Figure \ref{fig:basketHeat}. It is evident that general baskets tend to have a higher value than the others. A larger number of basket clusters leads to the creation of sparsely populated basket clusters with small product categories. Furthermore, these clusters vary depending on the season, which could cause problems in subsequent interpretation of the customer migration between segments.

\begin{table}
\begin{center}
\begin{tabular}{llr}
\toprule
Cluster & Description & Share \\
\midrule
B01 & General -- Big                           & 13\% \\
B02 & General -- Small                         & 14\% \\
B03 & Specialized -- Detergents                & 10\% \\
B04 & Specialized -- Laundry detergents        &  8\% \\
B05 & Specialized -- Body products             &  7\% \\
B06 & Specialized -- Face products             &  6\% \\
B07 & Specialized -- Dental products           &  9\% \\
B08 & Specialized -- Hair products             & 11\% \\
B09 & Specialized -- Beauty products           & 10\% \\
B10 & Specialized -- Products for men          &  2\% \\
B11 & Specialized -- Products for children     &  6\% \\
B12 & Specialized -- Feminine hygiene          &  5\% \\
\bottomrule
\end{tabular}
\caption{Distribution of the SM basket clustering.}
\label{tab:basketCount}
\end{center}
\end{table}

In the second step, we determine customer segments based on the ratio of basket archetypes they bought. We do not use the absolute number of baskets for purely practical reasons. Our goal is to estimate the shopping mission of ordinary customers. However, some people visit the retail chain to supply their own small business. They buy an enormous number of products with a huge frequency. Clustering algorithms, in that case, are likely to create numerous clusters just for a very small number of customers. This is a logical and right way. However, this information about the value and frequency is already included in the RFM segmentation. Therefore we standardize the number of baskets by using the ratio of basket archetypes bought by a customer. Customers with unusual shopping behavior are easily detectable using the RFM and SM segmentations as a whole, so we do not need to exclude them at all. For customer $C_i$, $i = 1, \ldots, n_C$, the SM segmentation is based on the information about basket cluster $j$, $j=1,\ldots,k_B$ given by
\begin{equation}
IC_i^{Bas,j} =  \frac{ \text{Number of baskets in $j$ purchased by $i$}}{ \text{Total number of baskets purchased by $i$}}.
\end{equation}
Customer $C_i$, $i = 1, \ldots, n_C$ is then assigned to the SM segment $C^{SM}_j$, $j = 1,\ldots,k_{SM}$ according to the vector variable
\begin{equation}
IC_i^{SM} = \left[ IC_i^{Bas,1}, \ldots, IC_i^{Bas,k_{B}} \right].
\end{equation}
For the second phase we also use the k-means algorithm and select the optimal number of clusters according to the Davies-Bouldin index and ratio of between cluster variance. 

\begin{figure}
\begin{center}
\includegraphics[width=0.7\textwidth]{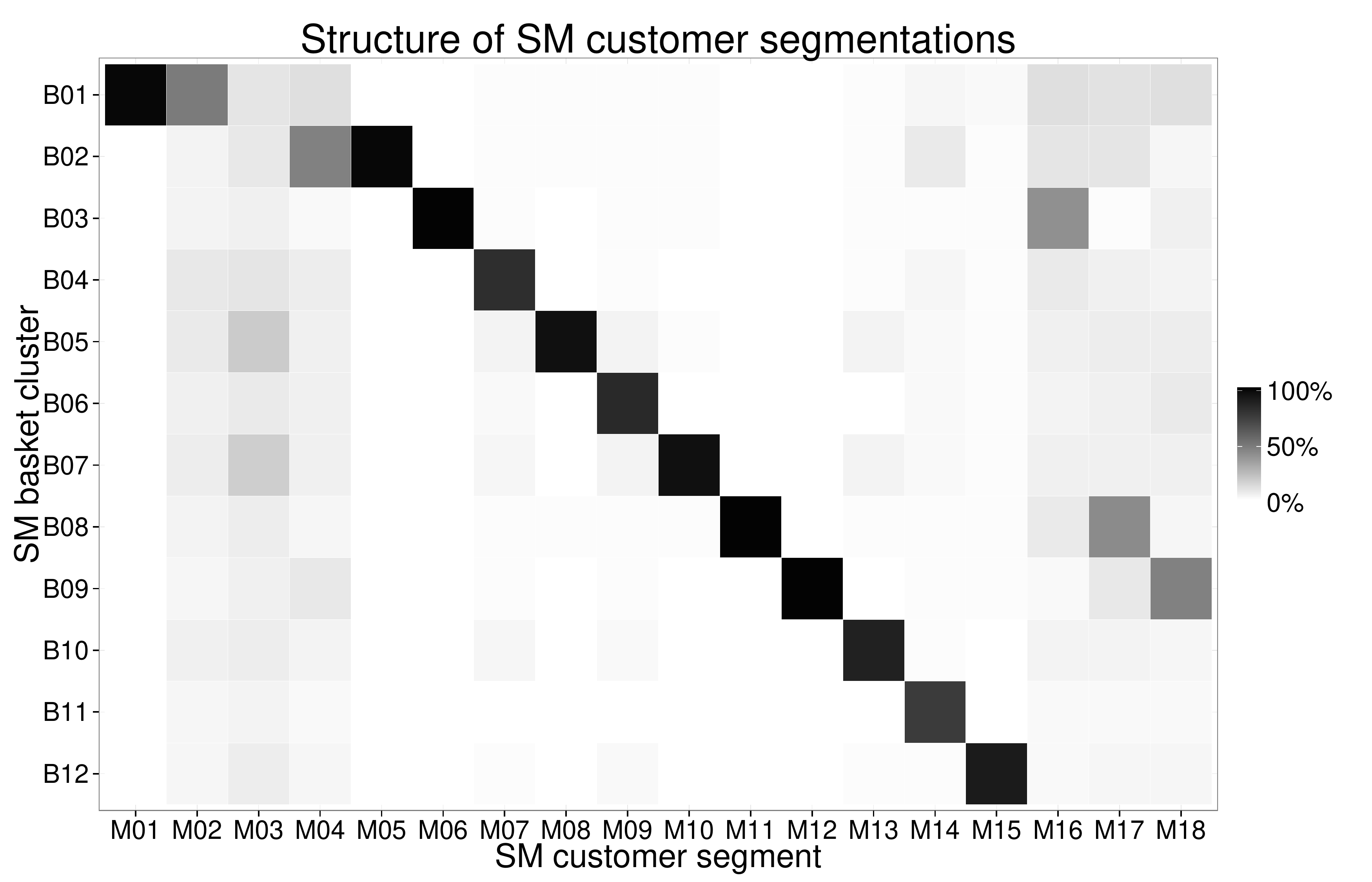} 
\caption{Ratios of baskets from the SM customer segments split into the SM basket clusters.}
\label{fig:missionHeat}
\end{center}
\end{figure}

We continue with our empirical analysis and segment customers of a Czech drugstore chain. We find the optimal number of clusters to be 18 using Davies-Bouldin index and between cluster variance ratio statistics. For the description of each segment, we use its center ratios of each basket type in the customer history. This allows us to distinguish three main types of customers. \textit{General} customers buy variety of categories in their baskets. As a customer with bulk purchases has a significantly different shopping motivation compared to a customer with very small yet varied purchases, the general group is further divided into more segments based on the prevailing purchase size. \textit{Focused} customers focus only on one type of category in each of their purchases and visit the store with a very straightforward motivation. They are looking for specific products and may be buying the rest of the drugstore products elsewhere. The proposed segmentation further divides focused customers into segments based on the category they prefer in majority of their purchases. The \textit{mixed} customers are a combination of the above customer types. Overall, the clustering consists of 5 general segments with different basket values, 10 focused segments formed around a single basket type and 3 mixed segments of both general and focused baskets. As long as the clusters are easy to interpret, a higher number of clusters is not a problem. The k-means method also provides a simple interpretation using centers of the clusters. As a result, managers without deep knowledge of the method can easily select groups of customers they want to reach out or analyze. The interpretation of the segments along with the percentage of assigned customers is shown in Table \ref{tab:missionCount}. The structure of the basket archetypes in clusters is shown in Figure \ref{fig:missionHeat}.

\begin{table}
\begin{center}
\begin{tabular}{llr}
\toprule
Cluster & Description & Share \\
\midrule
M01 & General -- Exclusively small           &  6\% \\
M02 & General -- Mainly small                &  9\% \\
M03 & General -- Small and big               &  8\% \\
M04 & General -- Mainly big                  & 11\% \\
M05 & General -- Exclusively big             &  7\% \\
M06 & Focused -- Detergents                  &  3\% \\
M07 & Focused -- Laundry detergents          &  4\% \\
M08 & Focused -- Body products               &  3\% \\
M09 & Focused -- Face products               &  4\% \\
M10 & Focused -- Dental products             &  3\% \\
M11 & Focused -- Hair products               &  4\% \\
M12 & Focused -- Beauty products             &  4\% \\
M13 & Focused -- Products for men            &  3\% \\
M14 & Focused -- Products for children       &  4\% \\
M15 & Focused -- Feminine hygiene            &  1\% \\
M16 & Mixed -- Detergents                    &  8\% \\
M17 & Mixed -- Hair products                 & 10\% \\
M18 & Mixed -- Beauty products               &  6\% \\
\bottomrule
\end{tabular}
\caption{Distribution of the SM customer segmentation.}
\label{tab:missionCount}
\end{center}
\end{table}

The division into the three main customer types and the subdivision into the specific segments is important in choosing suitable marketing strategies and is not contained in the common RFM and PPS segmentation techniques.  Using the knowledge of experts in the field and the analysis of customer characteristics, each segment can be further described. For example, the segment of customers focused exclusively on the big baskets is distinguished by high proportion of promotion sales. Not surprisingly, the segment of customers focused on products for children consists mainly of parents in their 20s and 30s. Such type of information is crucial for practical applications including marketing targeting and optimization of promotion sales. The proposed approach therefore offers a novel insight into the shopping behavior of customers.

Overall, our approach is similar to \cite{Reutterer2006} but there are some distinctive differences. Their first phase consists of assigning shopping baskets to the basket prototypes. This is done using clustering techniques applied to incidence matrix of purchased products leading to a high number of basket prototypes. We also need to assign shopping baskets to the clusters but instead of incidence matrix we use the percentage of total spending by product category along with the standardized total value of the basket. This allows us to consider lower number of clusters. The main difference between the methods, however, lies in the second phase. \cite{Reutterer2006} simply count occurrences of each customer's basket prototype during a given period. Customers are then assigned to behaviorally persisting segment if the number of purchased baskets of some type exceed a user-defined threshold value. Instead, we use k-means for the ratios of purchased basket types for each customer. The approach of \cite{Reutterer2006} is intended to be used in longer time-frames and utilizes series of historical data. Our method does not utilize the historical data which may be an advantage.

\section{Comparison of Segmentations}
\label{sec:combination}

First, we compare the RFM, PPS and SM segmentations. Our goal is to find if the segmentations are similar or if each segmentation brings unique information to the customer analysis. We adopt the \emph{purity} measure for comparison. Let us assume we have $n$ objects clustered by methods $I$ and $II$ with $k_{I}$ and $k_{II}$ clusters. The purity is then defined as
\begin{equation}
\text{Purity} = \frac{1}{n} \sum_{i=1}^{k_I} \max_{j} | C^I_i \cap C^{II}_j |,
\end{equation}
where $C^{I}_i$ is the set of objects in the cluster $i$ of the method $I$ and $C^{II}_j$ is the set of objects in the cluster $j$ of the method $II$. Therefore, the purity is a measure of the extent to which the clusters from clustering $I$ contain a single cluster from clustering ${II}$. Similar clusterings have the purity close to 1 while unsimilar clusterings have the purity close to 0. Note that the purity is not symmetrical. Table \ref{tab:purity} reports the purity for segmentations based on recency (Rec), frequency (Frq), monetary value (Mon), purchased product structure (PPS) and shopping mission (SM). We can see that each segmentation is unique as there are no segmentations with a high similarity. However, a medium similarity does exist. For example, the F and PPS segmentations are related to the SM approach due to Frq/SM purity $0.52$ and PPS/SM purity $0.43$. Reverse relationships have much lower purities because the SM segmentation has significantly more clusters than other segmentations.

\begin{table}
\begin{center}
\begin{tabular}{lrrrrr}
\toprule
      &   Rec &   Frq &   Mon &   PPS &    SM \\
\midrule
Rec   & 1.00 & 0.30 & 0.21 & 0.19 & 0.27 \\
Frq   & 0.48 & 1.00 & 0.40 & 0.32 & 0.52 \\
Mon   & 0.33 & 0.37 & 1.00 & 0.27 & 0.31 \\
PPS   & 0.43 & 0.42 & 0.42 & 1.00 & 0.43 \\
SM    & 0.18 & 0.18 & 0.13 & 0.13 & 1.00 \\
\bottomrule
\end{tabular}
\caption{Purity between customer segmentations in rows and columns.}
\label{tab:purity}
\end{center}
\end{table}

\begin{figure}
\begin{center}
\includegraphics[width=0.7\textwidth]{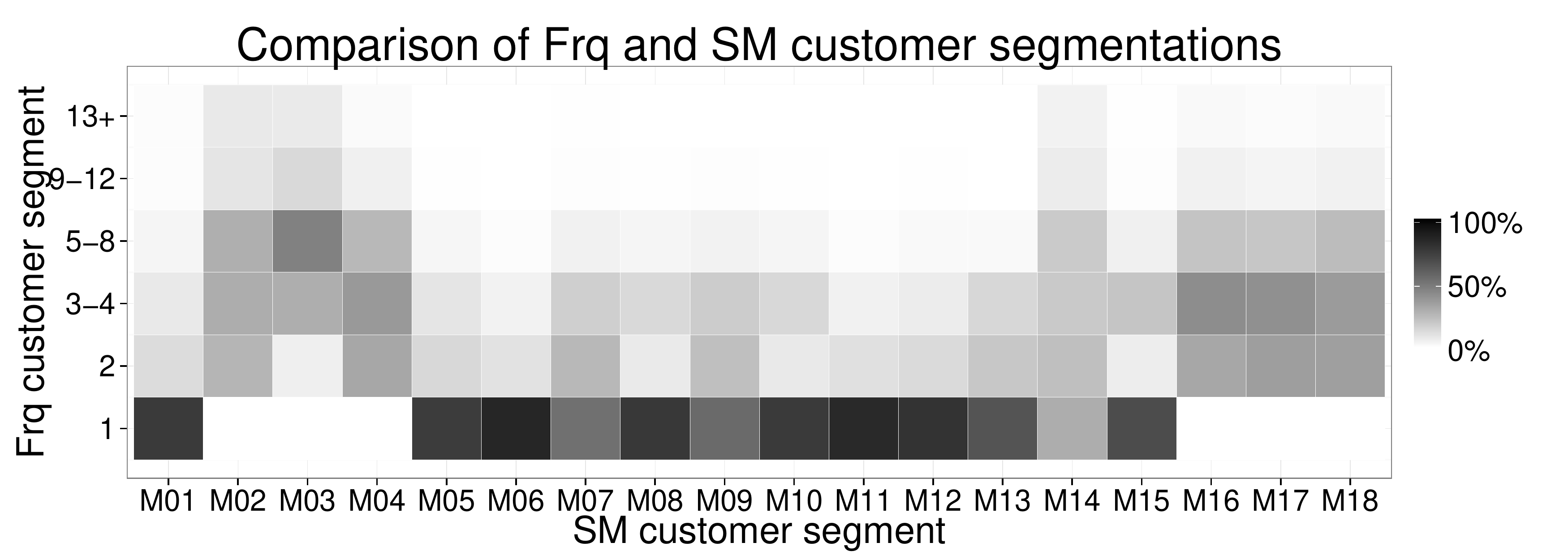} 
\caption{Ratios of customers from the SM segments split into the Frq segments.}
\label{fig:frqSmHeat}
\end{center}
\end{figure}

Next, we investigate the relationship between the SM and Frq segmentations in more detail. We compare the Frq and SM segments described in Table \ref{tab:missionCount}. Figure \ref{fig:frqSmHeat} shows how customer segments from the SM approach are divided into the Frq segments. The general segments M02--M04 and mixed segments M16--M18 have relatively high frequency while general segments M01 and M05 and the specialized focused segments M05--M15 have quite low frequency. This is an expected result as loyal customers with a general shopping visit the shop more often than customers that mainly shop elsewhere and visit the shop only for emergencies.

\begin{figure}
\begin{center}
\includegraphics[width=0.7\textwidth]{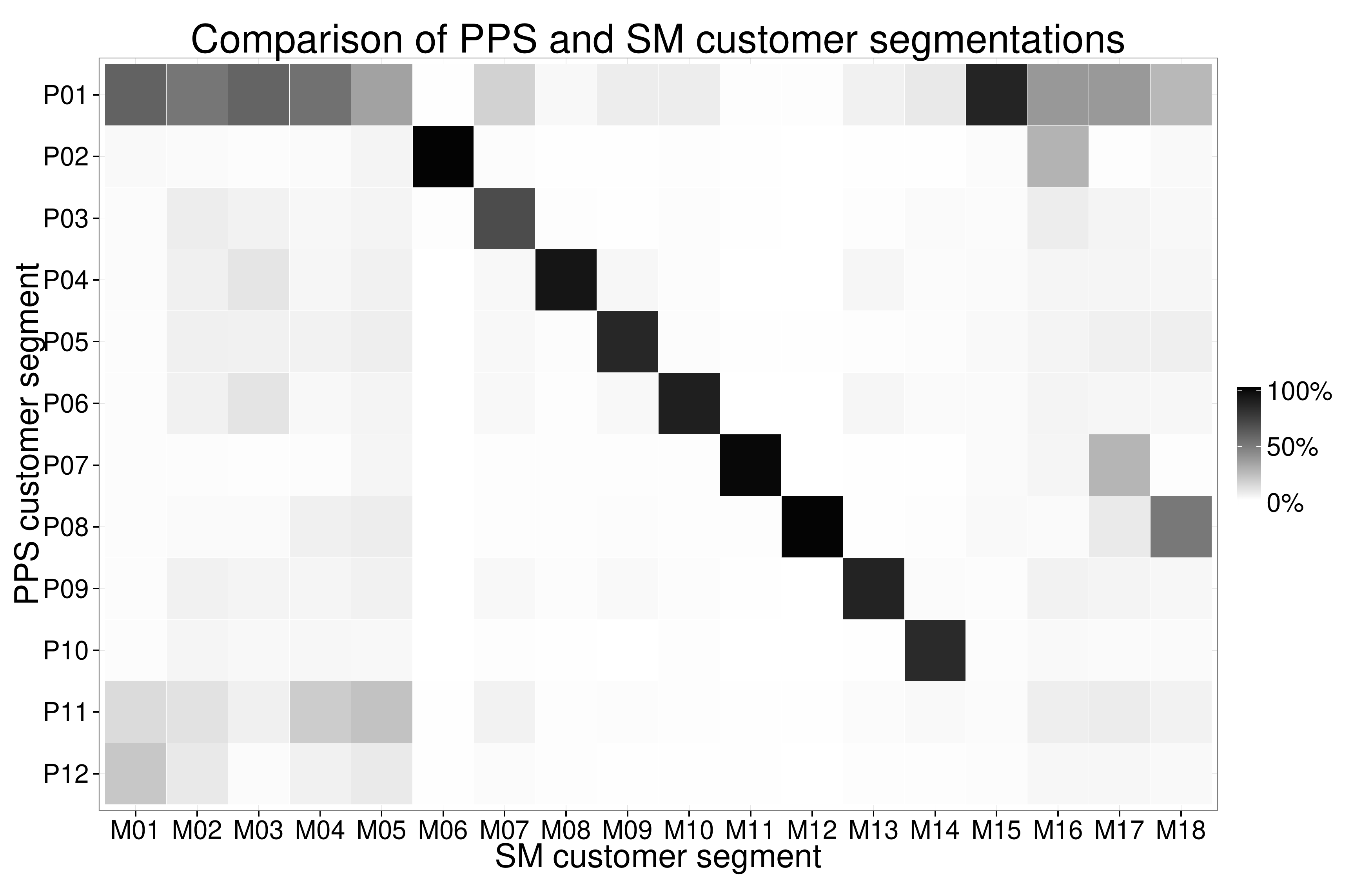} 
\caption{Ratios of customers from the SM segments split into the PPS segments.}
\label{fig:ppsSmHeat}
\end{center}
\end{figure}

Finally, we investigate the relationship between the SM and PPS segmentations in more detail. We compare PPS segments described in Table \ref{tab:productCount} and SM segments described in Table \ref{tab:missionCount}. Figure \ref{fig:ppsSmHeat} shows how customer segments from the SM approach are divided into the PPS segments. We can see that specialized segments P02--P10 correspond to segments M05--M14. General segment P01 is split among M01--M05 clusters according to the value of typical baskets and to clusters M16--M18 with dominant products. Interestingly, feminine hygiene products form their own segment M15. This is because many customers visit the drugstore specifically for these products but also purchase them in general baskets. Specialized segments P11 and P12 are clustered into general segments M01--M05. This is because customers do not visit shop specifically for these products but rather buy them together with other products.

All considered segmentations should be used together in the analysis of customers as each segmentation has a unique structure and interpretation and brings different information about customers.

\section{Implications for Practice}
\label{sec:practice}

Finally, we discuss the application of the proposed approach to a Czech drugstore. The main use of the proposed segmentation lies in better planning of promotional activities and targeted communication. Based on the proposed customer segmentation, the communicated message can be modified into a few variants according to the customer's interest. A major advantage of this approach is its simple interpretation as managers can easily select groups of customers based on their shopping mission. The typical situation in the e-mail targeting is the following. Every two weeks dozens of products are discounted and promoted. In e-mail communication, however, only a subset of them are announced. In order to maximize the effect of such promotions, it is necessary to select the products that will attract the customer the most. In our experience, only a few variations of communicated message to customers have good results and are the most cost-effective approach.

In order to examine the effect of the proposed segmentation, we conducted the following experiment. We divided customers from each segment randomly into test and control groups. The standard e-mail was sent to the customers in the control group while targeted e-mails designed according to the customer segmentation were sent to the customers in the test group. We focused mainly on three large groups of customers -- customers focused on products for children (segment \textit{M14}), customers focused on detergents (segments \textit{M06, M07} and \textit{M16}) and customers focused on hair and beauty products (segments \textit{M11, M12, M17} and \textit{M18}). After the implementation of the SM segmentation, we observed an increase in the opening of e-mails for all test groups in comparison to control groups. The increase was 11 percent for the children group, 3 percent for the detergents group and 2 percent for the hair and beauty group. We also observed an increase in average spending in the promoted products by customers who opened the e-mail. The difference was 6 percent for the children group, 2 percent for the detergents group and 5 percent for the hair and beauty group.

Furthermore, we compared the newly introduced SM segmentation in the current period with the PPS segmentation used by the managers in the previous period. In order to avoid bias due to irrelevant changes between the two periods, we standardized the open rate and spending indicators according to the average performance of the total e-mail wave. We observed an increase in the open rate in all three groups. The increase was less than one percent for the children group, 4 percent for the detergents group and 4 percent for the hair and beauty group. Negligible increase in the test group of customers focused on children products was caused by the high degree of similarity between the PPS and SM segments focused on children. Next, we focused on the average spending on the promoted products. Similar to the open rate, the children products group remained at the same level. In contrast, the spending increased by 3 percent in detergents group and by 6 percent in the hair and beauty group. Overall, the proposed SM segmentation brought the highest effect for premium groups focused on hair and beauty products. The effect on customers focused on detergents, which is generally less premium category, was smaller. The difference between the customer group focused on children products was minimal compared to the previous period due to the fact that the PPS and SM segments are very similar for customers with children.

\section{Conclusion}
\label{sec:conclusion}

We deal with a segmentation of customers in retail business according to their shopping behavior. The paper has two main contributions.
\begin{itemize}
\item First, we propose a new segmentation approach based on a shopping mission of a customer. The shopping mission answers the question why the customer visits the shop. Possible shopping missions include the purchase of a specific product category and the general purchase.
\item Second, we show how various segmentations can be combined in a real application. Besides the proposed method, we also consider recency, frequency and monetary value approach as well as the approach based on the structure of purchased products. The results show that the proposed segmentation brings useful insight into the analysis of customer behavior.
\end{itemize}
The proposed segmentation has been introduced in a major Czech drugstore chain and is currently used mainly for the e-mail targeting. The customer reaction indicators in targeted emailing campaigns such as the open rate and click rate have significantly improved in comparison to the previously used PPS segmentation.

\section*{Acknowledgements}
\label{sec:acknow}

We would like to thank Michal Černý for his comments and Alena Holá with Zuzana Veselá for proofreading.

\section*{Funding}
\label{sec:fund}

This work was supported by the Czech Science Foundation under Grant 20-17529S and the Internal Grant Agency of the University of Economics, Prague under Grant F4/21/2018.


\end{document}